\documentclass[aps,prb,twocolumn]{revtex4}
\usepackage{graphicx}
\usepackage{bm}

\newcommand{\be}{\begin{equation}}
\newcommand{\ee}{\end{equation}}
\newcommand{\beq}{\begin{eqnarray}}
\newcommand{\eeq}{\end{eqnarray}}

\begin{document}

\title{Vortex pinning by a columnar defect in planar superconductors
with point disorder}
\author{Anatoli Polkovnikov, Yariv Kafri and David~R.~Nelson}
\address{Physics Department, Harvard University, Cambridge, MA 02138}
\date{\today}

\begin{abstract}
We study the effect of a single columnar pin on a
$(1+1)$-dimensional array of vortex lines in planar type-II
superconductors in the presence of point disorder. In large
samples, the pinning is most effective right at the temperature of
the vortex glass transition. In particular, there is a pronounced
maximum in the number of vortices which are prevented from tilting
by the columnar defect in a weak transverse magnetic field. Using
renormalization group techniques we show that the columnar pin is
irrelevant at long length scales both above and below the
transition, but due to very different mechanisms. This behavior
differs from the disorder-free case, where the pin is relevant in
the low-temperature phase. Solutions of the renormalization
equations in the different regimes allow a discussion of the
crossover between the pure and disordered cases. We also compute
density oscillations around the columnar pin and the response of
these oscillations to a weak transverse magnetic field.
\end{abstract}

\maketitle

\section{Introduction}


A key challenge in the physics of vortex line arrays in
high-temperature superconductors (HTSC's) is understanding the
interplay between vortex interactions and various types of
pinning~\cite{blatter,natterman1}. The competition between thermal
fluctuations and pinning can lead to different phases such as
vortex liquids, Bose and vortex glasses, and a more ordered Bragg
glass~\cite{nv,devereaux,fisher_glass,ffh, nat_sch}.

Considerable theoretical progress can be made in studying
two-dimensional superconductors with an in-plane magnetic field,
where the vortex lines form one dimensional arrays. Experimentally
this situation can be realized using thin platelet superconducting
samples~\cite{bolle}. The statistical mechanics of such systems is
equivalent to the physics of interacting bosons in one dimension.
The low-energy and long wavelength properties can then be
described within a Luttinger-liquid formalism. For example, one
can relate the Bose-glass to vortex-liquid phase transition in the
presence of disordered columnar defects~\cite{nv} to the
superfluid-insulator transition in a system of interacting bosons
with quenched disorder~\cite{gs}. One finds that for a given
disorder strength the system can be tuned across the phase
transition by changing the temperature (which is proportional to
$g$, the Luttinger-liquid parameter). The mapping is different
with point disorder, which is equivalent to time-dependent point
impurities in the boson problem. In this case there is a subtle
second order phase transition between a ``supersolid'' (with
algebraic order both in boson and translational order parameter)
and glassy phase~\cite{Cardy,fisher_glass, hnv, Toner, hwa} with
decreasing temperature.

Another important feature of vortex physics in (1+1) dimensions is
the remarkable response to a single columnar defect. As argued,
originally in the quantum-mechanical context~\cite{kanefisher} and
later for vortex arrays, even a very weak columnar pinning
potential can grow to infinity under renormalization group (RG)
transformations when $g<1$. It was shown that the relevance of a
single columnar pin at low temperatures leads to a strong
suppression of a vortex tilt induced by a weak transverse magnetic
field. However, at high temperatures (such that $g>1$) the pin is
less effective, regardless of its microscopic strength.
Remarkably, the onset of the relevance (or irrelevance) of a
single columnar pin and point disorder occurs at the {\em same}
temperature $T^\star$, such that $g(T^\star)=1$.

\begin{figure}[ht]
\includegraphics[width=8.5cm]{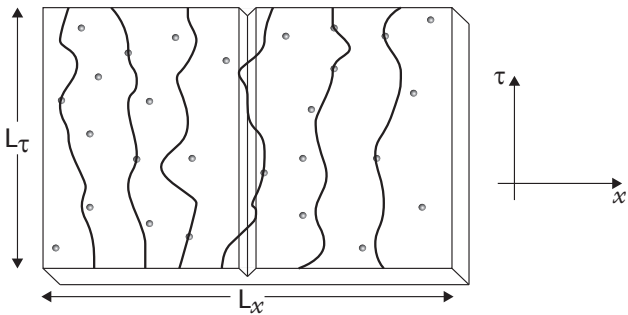}
\caption{Schematic view of a planar superconductor with point
disorder and a columnar pin, represented by a notch cut into the
sample. The wiggly lines correspond to vortices, which
alternatively can be thought of as imaginary time world lines of
bosons in one spatial dimension.}
\label{fig0}
\end{figure}

If both point disorder and a columnar pin are present (see
Fig.~\ref{fig0}) then at low temperatures ($g<1$) one expects a
competition between the two: a growing columnar pin strength under
renormalization leads to stronger correlated pinning of vortex
lines at long length scales. On the other hand, the increasing
point disorder tends to destroy the effect of the pin on distant
regions. Although it was argued that point disorder would always
render a single columnar defect irrelevant at long wavelengths in
Ref.~[\onlinecite{ahns1}], the precise nature of this competition
and the different pinning properties above and below the vortex
glass transition were left unresolved. A detailed study of this
competition is a primary goal of the present paper. We show that
in the thermodynamic limit the pinning strength is strongest
precisely at the transition point  $T^\star$ where $g=1$. In
particular, the number of pinned vortex lines is a nonmonotonic
function of $g$ and strongly peaked at $g=1$. For finite systems
the position of the maximum is slightly shifted to lower values of
$g$ (see Fig.~\ref{fig3a}). We emphasize that irrelevance of a
single columnar pin does not imply irrelevance of an array of
pins. In fact, as it was argued in
Refs.~[\onlinecite{nv,devereaux}], the vortex glass phase formed
by point disorder is unstable towards infinitesimal disorder in
columnar pins, which results in a Bose glass phase.

The paper is organized as follows. In Sec.~\ref{sec:model} we
introduce the model and describe the mapping to a Luttinger
liquid. In Sec.~\ref{sec:RG} the renormalization equations
describing the long wavelength physics are derived. Then in
Sec.~\ref{sec:FR} we analyze density oscillations of vortex lines
near the columnar pin above and below the vortex glass transition.
At the transition point $g=1$, we demonstrate that the problem
maps onto free Fermions in a time-dependent disorder potential.
This mapping allows computation of density oscillations of vortex
lines near the columnar pin which we then compare with the
renormalization group predictions. In Sec.~\ref{sec:tilt} we
discuss the response of vortex lines to a weak transverse magnetic
field in the presence of a single columnar pin. Finally in
Sec.~\ref{conclusions} we summarize our results and present
conclusions. Appendix~\ref{sec:ap_a} estimates the parameter range
necessary to see the effects discussed here, while
Appendix~\ref{sec:ap_b} describes details of our free fermion
calculation.

\section{Model}
\label{sec:model}

A one-dimensional array of vortex lines located at positions
$x_j(\tau)$ can be described by the density profile (see
Fig.~\ref{fig0})
\be
n(x,\tau)=\sum_j \delta[x-x_j(\tau)]\,.
\label{nxt}
\ee
Here $x$ and $\tau$ denote transverse and longitudinal coordinates
with respect to the vortices. It is convenient to change variables
to the phonon displacement field $u_j$:
$x_j(\tau)=a[j+u_j(\tau)]$, where $a$ is the mean distance between
the vortex lines. In the absence of a columnar pin, point
disorder, and a transverse magnetic field, the free energy of a
particular vortex line configuration can then be written
as~\cite{ahns1}
\be
\mathcal F_0={a^2\over 2}\int dxd\tau \; \left[c_{11}(\partial_x
u)^2+c_{44}(\partial_\tau u)^2\right],
\ee
where $c_{11}$ and $c_{44}$ are the compressional and the tilt
modulii, respectively. After rescaling  $x$ and $\tau$
\be
x\to x \left(c_{11}\over c_{44}\right)^{1/4},\; \tau \to \tau
\left(c_{44}\over c_{11}\right)^{1/4},
\label{resc}
\ee
the free energy takes the isotropic form
\be
\mathcal F_0={A\over 2}\int dxd\tau \; \left[(\partial_x
u)^2+(\partial_\tau u)^2\right]\,,
\label{f0}
\ee
with $A=a^2\sqrt{c_{11}c_{44}}$. The partition function $Z$
describing a vortex array at temperature $T$ is a functional
integral over all possible configurations of vortices weighted by
a Boltzmann factor proportional to $e^{-\mathcal F_0/T}$.
In the limit of large sample dimension in a ``timelike''
direction, $Z$ can also be regarded as the zero temperature
partition function of interacting bosons~\cite{nv,ns}
\be
Z=\int Du(x,\tau)\mathrm e^{-S},
\label{zq}
\ee
with the imaginary time action $S=S_0=\mathcal F_0/T$ given by
\be
S_0={\pi\over 2g}\int dxd\tau \left((\partial_x
u)^2+(\partial_\tau u)^2\right).
\label{act}
\ee
For simplicity we here set the Planck's constant $\hbar=1$.
Comparison of Eq.~(\ref{act}) with Eq.~(\ref{f0}) allows us to
identify the Luttinger-liquid parameter $g$ as
\begin{displaymath}
g={\pi T\over A}.
\end{displaymath}
The rescalings in Eq.~(\ref{resc}) are such that the ``sound

The most relevant contributions to the action from the columnar
pin $S_{pin}$ and point disorder $S_{pd}$ read~\cite{fisher_glass,
hnv, hwa, AHNS}
\be
S_{pin}=V_0\int d\tau \cos[2\pi u(0,\tau)],
\label{sp1}
\ee
\be
S_{pd}=2\int dxd\tau U(x,\tau) \cos[2\pi
u(x,\tau)+\beta(x,\tau)]\,,
\label{spd1}
\ee
where positive (negative) $V$ and $U$ correspond to repulsive
(attractive) potentials. When $V<0$, Eq.~(\ref{sp1}) represents an
attractive columnar pin at the origin. For simplicity we take the
phase $\beta(x,\tau)$ to be uniformly distributed between $0$ and
$2\pi$ and $U(x,\tau)$ to have a Gaussian distribution with the
correlator
\be
\overline{U(x_1,\tau_1) U(x_2,\tau_2)}=\Delta_0
\delta(x_1-x_2)\delta(\tau_1-\tau_2),
\label{delta}
\ee
where the overbar represents an average over realizations of the
disorder. The total action $S$ entering Eq.~(\ref{zq}) is then the
sum of the three contributions (\ref{act}), (\ref{sp1}), and
(\ref{spd1}): \be S=S_0+S_{pin}+S_{pd}. \label{act1} \ee In the
following sections we analyze $S$ and various observables using
the renormalization group. Before proceeding with quantitative
details we emphasize that although we focus on the behavior of
vortex lines in this paper, the action (\ref{act1}) can be
relevant for many different problems such as disordered
interfaces~\cite{Toner}, charge density waves which order like
smectic liquid crystals,~\cite{glatz, bellini} and directed
polymer arrays~\cite{vilgis}.

\section{Renormalization group flow equations}
\label{sec:RG}

Provided one is alert to potential patologies~\cite{fisher}, an
efficient way to analyze disordered problems is to use a replica
trick~\cite{anderson}. The noninteracting part of the action
(\ref{act}) then becomes
\be
S_0={\pi\over 2g}\sum_{\alpha,\beta} \int\int dxd\tau
\bigg[{\partial u_\alpha\over \partial\tau}{\partial u_\beta\over
\partial\tau}+{\partial u_\alpha\over \partial
x}{\partial u_\beta\over
\partial x}\bigg]\bigg[\delta_{\alpha\beta}-{\kappa\over  g}\bigg],
\label{1}
\ee
where $u_\alpha(x,\tau)$ is the replicated phonon field and
$\kappa$ is an off-diagonal coupling which is zero in the bare
model but generated by the disorder~\cite{Cardy}. It is equivalent
to ($x$- and $\tau$-dependent) quenched random ``chemical
potential'' coupled to the first derivatives of the phonon field
$u$~\cite{ahns1, hwa}. The replica indices $\alpha$ and $\beta$
run from $1$ to $n$ and we take the limit $n\to 0$ at the end of
the calculation~\cite{Cardy,Toner}. Equation~(\ref{1}) leads to a
phonon correlation function in momentum space for $n\to 0$,
namely,~\cite{Cardy}
\be
\langle
u_\alpha(k,\omega)u_\beta^\star(k^\prime,\omega^\prime)\rangle={4\pi
\over k^2+\omega^2}\left(g\delta_{\alpha\beta}+\kappa
\right)\delta_{k,k^\prime}\delta_{\omega,\omega^\prime}.
\ee
The other two terms in the action of Eq.~(\ref{act1})
corresponding to a pinning potential and point disorder, become
\be
S_{pin}=V_0 \sum_\alpha\int d\tau \cos\, 2\pi u_{\alpha}(0,\tau)
\label{spin}
\ee
and
\be
S_{pd}=-\Delta_0\sum_{\alpha,\beta}\int\int dxd\tau\cos\,
2\pi\big[u_{\alpha}(x,\tau)-u_{\beta}(x,\tau)\big].
\label{spd}
\ee

To study the statistical physics described by the action
(\ref{1}), (\ref{spin}), and (\ref{spd}) we employ a momentum
shell renormalization group scheme~\cite{Cardy1}, where we
continuously eliminate degrees of freedom depending on frequency
and momentum within the shell
$\Lambda-\delta\Lambda<\sqrt{\omega^2+k^2}<\Lambda$. Here
$\Lambda\sim 1/\sqrt{a_0\xi_0}$ is the ultraviolet cutoff, $a_0$
is of the order of the lattice spacing, and $\xi_0$ is of the
order of superconducting coherence length (single vortex width).
The resulting renormalization group equations for the running
coupling constants $\Delta(l)$ and $V(l)$ to leading order in
$\Delta$ and $V$ are \beq
&&{dg\over d l}=0,\label{3c}\\
&&{d\Delta\over d l}=2\epsilon\Delta-2C\Delta^2,\label{3a}\\
&&{d\kappa\over d l}=C^2\Delta^2,
\label{3b}\\
&& {dV\over dl}=V\left(\epsilon-C\Delta\right)-\kappa V,
\label{3d}
\eeq
where $\epsilon=1-g$, $l$ is the flow parameter
($\Lambda(l)=\Lambda\mathrm e^{-l}$), and $C$ is a nonuniversal
constant which depends on the cutoff $\Lambda$: $C\propto
1/\Lambda^2$ (in particular, within the shell method we find
$C=8\pi g^2/\Lambda^2$). These equations are subject to the
initial conditions $\kappa(l=0)=0$, $\Delta(l=0)=\Delta_0$, and
$V(l=0)=V_0$, with $\Delta_0$ and $V_0$ being the bare couplings.
Note that the Luttinger-liquid parameter $g$ does not change under
renormalization~\cite{Cardy}.

In the absence of point disorder ($\Delta(l)\equiv 0$ and
$\kappa(l)\equiv 0$) our results reduce to those obtained by Kane
and Fisher~\cite{kanefisher}. In this case the columnar pin is
relevant for $g<1$ and irrelevant for $g>1$. If $V\equiv 0$ then
our equations are equivalent to those first derived by Cardy and
Ostlund~\cite{Cardy} and later extensively explored for different
problems~\cite{Toner, hwa, fisher_glass, hnv}. Equations
(\ref{3c})-(\ref{3b}) imply that point disorder {\em also} becomes
relevant when $g<1$. Contrary to bosons interacting with many
impurities~\cite{gs} (which is the analogue of many {\em columnar}
defects for a flux problem), there is an intermediate fixed point
with a finite value $\Delta^\star\equiv\lim_{l\to
\infty}\Delta(l)=O(\epsilon)$ which continuously emerges from a
pure Gaussian fixed point for $g<1$. This makes the RG approach
tractable on both sides of $g=1$. Note also that when $g<1$,
$\kappa(l)\to\infty$, suggesting nontrivial correlations in this
phase. We comment that there are some claims questioning the
applicability of Eqs.~(\ref{3c})-(\ref{3b}) in the glass phase
($g<1$). In Ref.~[\onlinecite{doussal_giamarchi}] it was argued
that a replica symmetric solution becomes unstable for $g<1$,
resulting in different correlation functions than predicted by the
replica symmetric renormalization group. However, there is still
no evidence showing that this instability actually occurs.
Moreover, numerical results of Ref.~[\onlinecite{zeng}] confirm
one of the crucial predictions of Eqs.~(\ref{3c})-(\ref{3b}) at
$g<1$; namely, the unusual behavior of the density-density
correlation function $G(x)\propto \exp(-A\ln^2|x|)$.

In deriving equations (\ref{3a})-(\ref{3d}), we implicitly assumed
that the cutoff is symmetric in the $\tau$ and $x$ directions. In
general this is not true. The anisotropy in the cutoff will result
in different initial renormalizations of $\kappa_\tau$ and
$\kappa_x$. However, at large length scales the flows for
$\kappa_\tau$ and $\kappa_x$ look the same and the asymmetry
disappears~\cite{hwa}.

Equations (\ref{3c})-(\ref{3d}) contain nonuniversal cutoff
dependent terms. Upon rescaling the disorder potential
$\Delta\to\tilde\Delta/C$ they simplify to
\beq
&& {dg\over dl}=0,\label{flow0}\\
&&{d\tilde\Delta\over d l}=2\tilde\Delta(\epsilon-\tilde\Delta),
\label{flow1}\\
&&{d V\over d l}=V(\epsilon -\tilde\Delta)-\kappa  V,
\label{flow2}\\
&&{d\kappa\over d l}=\tilde\Delta^2. \label{flow3} \eeq These are
the renormalization equations which we will exploit throughout the
rest of the paper. Note that the cutoff $\Lambda$ enters
Eqs.~(\ref{flow0})-(\ref{flow3}) only through the initial disorder
strength $\tilde\Delta_0$.

The quantitative predictions of the renormalization group
equations above are valid if $\tilde\Delta_0$ is small compared to
one. On the other hand if the point disorder is too weak, then its
effects will be hard to observe in experiments. In
Appendix~\ref{sec:ap_a} we show that for HTSC superconducting
films, typical values of point disorder strength lie within
interval $\tilde\Delta_0\in [0.01, 0.1]$. This, in turn, implies
experimental relevance of the subsequent analysis of the physics
resulting from Eqs. (\ref{flow0})-(\ref{flow3}) in the different
regimes.

\subsubsection*{High temperature phase ($g>1$)}

If $\epsilon$ is large and negative ($|\epsilon|\gg \Delta_0$)
then both point disorder and pinning strengths decay exponentially
to zero as
\beq
&&\tilde\Delta(l)\sim \tilde\Delta_0\mathrm e^{-2|\epsilon| l}\to
0,
\label{tilde_delta}\\
&&V(l)\sim V_0\mathrm e^{-|\epsilon| l}\to 0.
\eeq
However, the off-diagonal stiffness $\kappa$ renormalizes to a
finite nonuniversal value
\be
\kappa(l)\approx {\tilde\Delta_0^2\over 4|\epsilon|}
\left(1-\mathrm e^{-4|\epsilon| l}\right)\to
{\tilde\Delta_0^2\over 4|\epsilon|}.
\ee
As we will see below, the finite value of $\kappa(l=\infty)$
(which arises for any $\epsilon\leq 0$) results in corrections to
the power law decay of various correlation functions.

\subsubsection*{Critical phase ($g=1$)}

When $\epsilon=0$ (i.e. $g=1$), point disorder becomes marginally
irrelevant, and one finds
\beq
&&\tilde\Delta(l)=\frac{\tilde\Delta_0}{1+2l\tilde\Delta_0}\to
0,\label{del}
\\
&&V(l)=\frac{V_0}{(1+2l\tilde\Delta_0)^{1/4}}\mathrm
e^{-{l\tilde\Delta_0\over 2}}\to 0,\\
&&\kappa(l)={\tilde\Delta_0-\tilde\Delta(l)\over 2}\to
{\tilde\Delta_0\over 2}.
\label{kap}
\eeq
Note that even though the point disorder $\tilde\Delta$ is
marginal at $g=1$, the pinning potential $V$ remains irrelevant
(i.e., $V(l)\to 0$) for any nonzero $\tilde\Delta_0$.

\subsubsection*{Low-temperature phase ($g<1$)}

In the case $\epsilon>0$ we find the following solutions of the
flow equations
\beq
&&\tilde\Delta(l)= {\epsilon\tilde\Delta_0\over
\tilde\Delta_0+(\epsilon-\tilde\Delta_0)\,\mathrm e^{-2\epsilon
l}}\to\epsilon,
\label{eps}\\
&&\kappa(l)= {\epsilon\over 2}\ln\left[1+{\tilde\Delta_0\over
\epsilon}\left(\mathrm e^{2\epsilon
l}-1\right)\right]+{\tilde\Delta_0-\tilde\Delta(l)\over
2}\nonumber\\
&&~~~~~~~~~~~~~~~~~~~~~~~~~~~~~~~~~~~~~~~~~\to \epsilon^2
l\to\infty.
\label{e2}
\eeq
Note that $\kappa(l)$ grows without bound. The explicit
mathematical expression for the renormalized columnar pinning
potential is rather complicated. However, one can write the
asymptotic form of $V(l)$ at large and small $l$:
\be
V(l)\approx \left\{\begin{array}{ll}
V_0\exp(\epsilon-\tilde\Delta_0) l, &
l\ll l_0,\\
V^\prime\exp\left[\left({\epsilon-\tilde\Delta_0\over
2}-{\epsilon\over
2}\ln{\epsilon\over\tilde\Delta_0}\right)l-\epsilon^2 l^2\right],
& l\gg l_0.
\end{array}\right.
\label{v}
\ee
Therefore as $l\to\infty$, $V(l)\to 0$ {\em faster} than
exponentially in $l$. Here, $l_0$ represents a crossover scale
\be
l_0\approx {1\over 2|\epsilon|}\ln{|\epsilon-\tilde\Delta_0|\over
\tilde\Delta_0}\label{l0}
\ee
and
\be
V^\prime= V_0\exp\left[{1\over 4}\ln{\epsilon\over
\tilde\Delta_0}-{1\over 4}\rm{Li}_2\left(1-{\epsilon\over
\tilde\Delta_0}\right)\right]\, ,
\ee
where $\rm{Li}_2(x)$ is the polylog function~\cite{Levin}. In the
limit $\epsilon\gg\tilde\Delta_0$ we can use the asymptotic
expansion for $\rm{Li}_2(x)$ and get
\be
V^\prime\approx cV_0\exp\left({1\over 4}\ln{\epsilon\over
\tilde\Delta_0}+{1\over 8}\ln^2{\epsilon\over
\tilde\Delta_0}\right),
\ee
where $c$ is a number of the order of 1. Note that even though a
term of the order of $\epsilon^2$ appears in Eq.~(\ref{v}), its
presence is justified. Indeed, according to Eq.~(\ref{flow3}), at
large $l$ we have $\kappa\propto \Delta^{\star\, 2}l\sim
\epsilon^2 l$. It is easy to see that higher order corrections in
$\epsilon$ to the renormalization group flow equations (in
particular to Eq.~(\ref{flow1})) will result in terms of the order
of $O(\epsilon^3)$ in Eq.~(\ref{e2}).

The parameter $l_0$ defined in Eq.~(\ref{l0}) sets a
characteristic length scale $\Lambda^{-1}\mathrm e^{l_0}$,
separating long and short length behavior of the pinning
potential. As we find below, it also determines the behavior of
various observables. Thus, for $\epsilon>0$, at small $l$ the
pinning potential first grows under the RG transformations to the
value $V_{max}\approx V(l_0)\approx
V_0\sqrt{\epsilon/\tilde\Delta_0}$. Then, for larger $l$, $V(l)$
goes to zero faster than exponentially.  We comment that for
$g>1$, $l_0$ sets the characteristic scale beyond which
$\Delta(l)$ becomes negligibly small and $\kappa(l)$ stops
renormalizing.

Note that the columnar pin is asymptotically irrelevant in the
presence of point disorder for all values of $g$. The mechanisms,
which lead to this are different below and above the vortex glass
transition. Thus in the high-temperature phase $g\geq 1$ thermal
fluctuations are responsible for the irrelevance of $V$ at
$l\to\infty$. In contrast, in the low-temperature glass phase
$g<1$ point disorder is the cause of the flow of $V$ to zero at
large $l$. The distance when the columnar pin starts feeling
effects of point disorder and becomes irrelevant grows with
decreasing $\tilde\Delta_0$. As discussed below, in infinite
samples, the effect of the columnar pin is strongest (least
irrelevant) precisely at $g=1$.

In the weak disorder limit one can compute various correlation
functions using the renormalization group analysis sketched above.
In what follows we will discuss several quantities of interest.

\section{Density oscillations and the free fermion limit}
\label{sec:FR}

\subsection{Density oscillations near a columnar pin}

Since the columnar pinning potential is always irrelevant when
point disorder is present, it can be treated perturbatively at
sufficiently large length scales. The leading contribution to the
``Friedel oscillations'' of the density of vortex lines in linear
response in $V_0$ is given by
\be
\overline{\langle \delta n(x)\rangle}\approx V_0 \cos 2\pi n_0
x\,\int_{-\infty}^\infty f(x,\tau) d\tau,
\label{34}
\ee
where the angular brackets represent the thermal average and the
overbar signifies an average over different configurations of
point disorder. The quantity
\be
\delta n(x)=n(x)-n_0
\ee
is the deviation of the vortex lines density from the mean
$n_0=1/a$ (see Eq.~(\ref{nxt})). The function $f(x,\tau)$ is
defined as
\be
f(x,\tau)=\overline{\langle \mathrm e^{2\pi i
[u(x,\tau)-u(0,0)]}\rangle}.
\ee
We note that $f(x,\tau)$ is proportional to the density density
correlation function without the columnar pin: $\langle\delta
n(x,\tau)\delta n(0,0)\rangle\propto \cos [2\pi
n_0(x)]\,f(x,\tau)$. Explicitly, one finds
\be
f(x,\tau)\approx \mathrm e^{-2\int_0^\infty d l
[g+\kappa(l)][1-J_0(r\mathrm e^{-l})]},
\label{eq:30}
\ee
where $r=\sqrt{x^2+\tau^2}\Lambda$ is the distance between the two
points measured in the units of the original cutoff $\Lambda$. The
Bessel function $J_0(x)$ appearing in Eq.~(\ref{eq:30}) and in
other formulas below is nonuniversal and depends on the actual
details of the cutoff procedure. Instead of $J_0(x)$ one can use
another cutoff function $\tilde J(x)$, e.g., a Gaussian, as long
as it satisfies general requirements $\tilde J(x\to 0)\to 1$ and
$\tilde J(x\to\infty)\to 0$. If the disorder is absent in
Eq.~(\ref{eq:30}), i.e., $\kappa\equiv 0$, we recover the
well-known result for the Luttinger liquid $f(x,\tau)\propto
r^{-2g}$. If $g\geq 1$ then at long distances $f$ is given more
generally by $f(x,\tau)\propto r^{-\eta}$~[\onlinecite{Toner}],
where
\be
\eta= 2[g+\kappa(\infty)].
\label{eq:32}
\ee
Thus, when point disorder is irrelevant, the exponent of the
correlation decay becomes non-universal. For $g<1$ the asymptotic
expression for $f$ becomes~\cite{Cardy, Toner} $f(x,\tau)\propto
\mathrm e^{-\epsilon^2 \ln^2 x}$.

Upon using Eqs.~(\ref{34}) and (\ref{eq:30}) we find that the
behavior of $\langle\delta n(x)\rangle$ at large distances for
$g\geq 1$ is
\be
\overline{\langle\delta n(x)\rangle}\propto  V_0 {\cos 2\pi n_0
x\over x^{\eta-1}}.
\label{36}
\ee
This equation is valid only for  $x\gtrsim \exp(l_0)$, where $l_0$
is given by Eq.~(\ref{l0}). For smaller $x$ the exponent $\eta$
changes with $x$. For $g<1$ the crossover is now from power law
decaying correlations of the form (\ref{36}) with $\eta\approx 2g$
for $1\ll x \ll \exp(l_0)$ to a faster decay
\be
\overline{\langle \delta n(x)\rangle}\propto V_0\mathrm
e^{-\epsilon^2 \ln(x)^2}
\label{tt}
\ee
in the opposite limit $x\gg \exp(l_0)$.

\subsection{Free fermions}

If $g=1$ it is well known that using the Jordan-Wigner
transformation bosons can be exactly mapped to spinless free
fermions~\cite{Sachdev_book}. The transformation also holds in the
presence of a columnar pin and point disorder. The columnar pin
and the point disorder correspond to static and random
time-dependent potentials, respectively. The time-dependent
Hamiltonian which describes the fermions then reads
\be
\mathcal H_f(x,\tau)=-{1\over
2m}{d^2\over dx^2}+U(x,\tau)+V_0(x),
\label{hf}
\ee
where $U(x,\tau)$ is a random potential satisfying
\be
\overline{ U(x_1,\tau_1) U(x_2,\tau_2)}=\Delta_0
\delta(x_1-x_2)\delta(\tau_1-\tau_2).
\label{uxt}
\ee
The mass $m$ (corresponding to the tilt modulus in the original
flux line problem) sets the Fermi velocity $v_f=k_f/m$ (where
$k_f$ is the Fermi momentum), which represents the sound velocity
in the original boson/vortex problem.

If the sample length $L_\tau$ in the timelike direction is large,
then the partition function of the vortex array is proportional to
an appropriate matrix element of the corresponding quantum
problem~\cite{hatano}
\be
Z=\langle G|  T_\tau \mathrm
e^{-\int_0^\infty d\tau\int dx \mathcal H_f(x,\tau)} |G\rangle,
\label{zq1}
\ee
where $T_\tau$ is the usual (imaginary) time-ordering symbol. This
expression is the quantum-mechanical expectation value of the
evolution operator calculated in its many-fermion ground state
$|G\rangle$, for a given realization of point disorder. If the
Hamiltonian $\mathcal H_f$ is time independent, then
Eq.~(\ref{zq1}) reduces to the zero-temperature quantum partition
function. For $N$ noninteracting fermions the ground state can be
written as a Slater determinant of the single particle states.
However, because the Hamiltonian $\mathcal H_f$ is time dependent,
the states forming the Slater determinant will not be the
eigenstates of $\mathcal H_f$. Instead, they will consist of the
$N$ largest eigenvalues of the {\it evolution operator} (see
Appendix~\ref{sec:ap_b} for further details). Once $|G\rangle$ is
known one can easily calculate various observables. Here we
consider the vortex line density
\be
\overline{\langle n(x)\rangle }=\overline{{1\over Z} \langle
G|c^\dagger (x)c(x)|G\rangle },
\ee
where $c(x)$ is a fermionic annihilation operator. Sufficiently
far from the boundaries at $\tau=0,L_\tau$, the density profile
$\overline{n(x)}$ clearly does not depend on $\tau$.

\begin{figure}[ht]
\includegraphics[width=8.3cm]{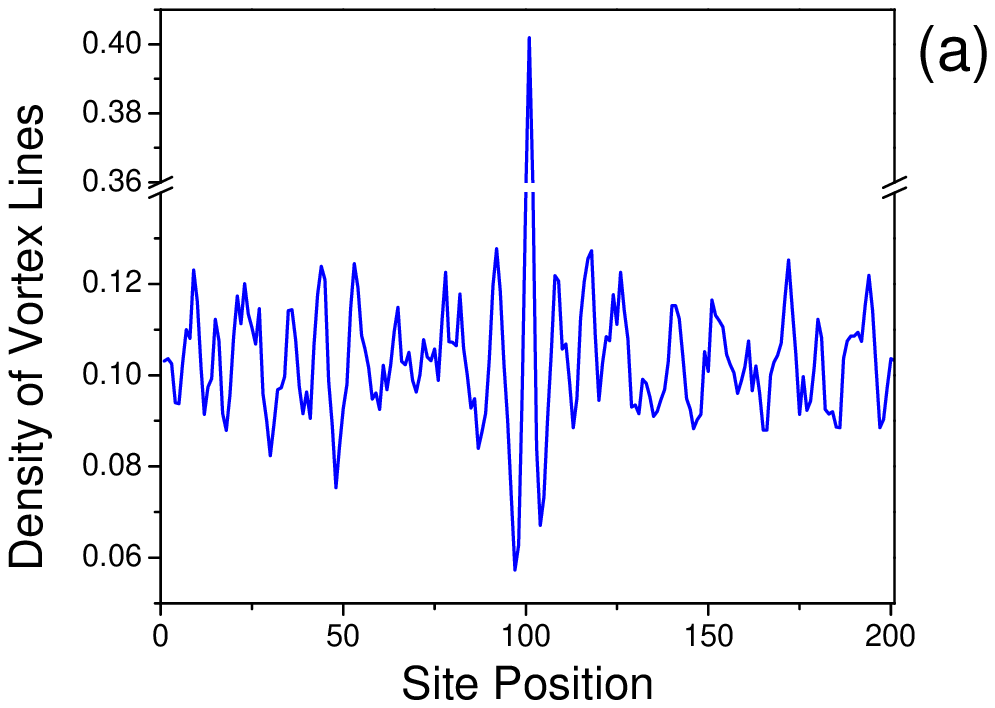}
\includegraphics[width=8.3cm]{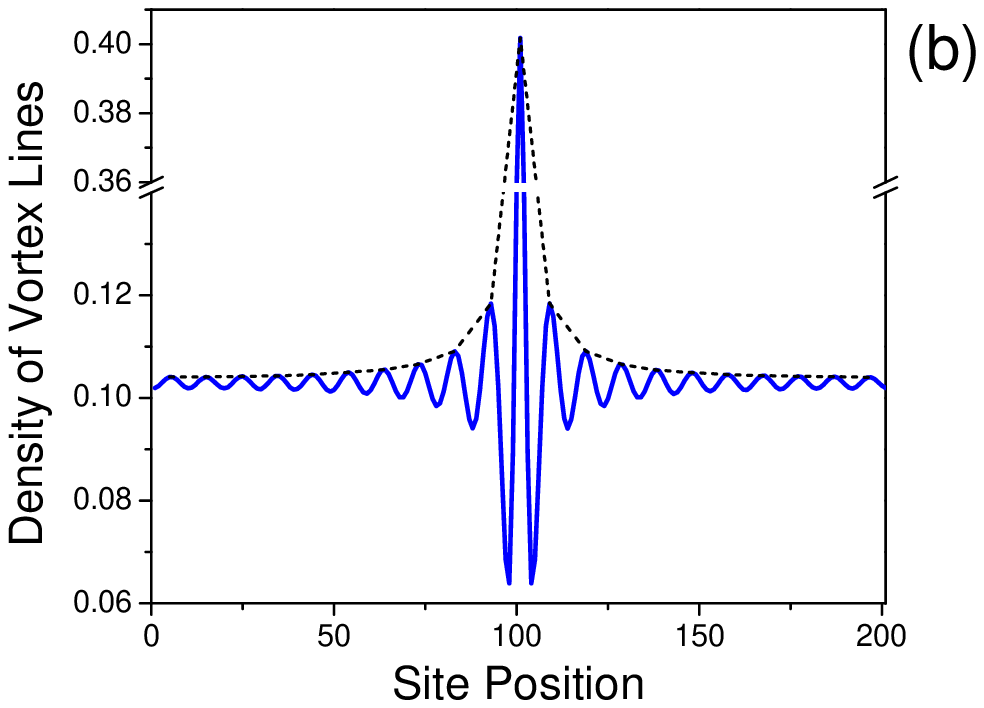}
\caption{Density profile of vortex line array near a columnar pin.
The top graph (a) shows the result for a single configuration of
point disorder. The bottom graph (b) gives the profile after
averaging over many disorder realization. The dashed line in the
bottom graph shows the envelope of the decay os oscillations,
which is used to extract the exponent $\eta$ (see
Eq.~(\ref{36})).}
\label{fig1a}
\end{figure}

Since we are dealing with noninteracting particles, one can find
the eigenstates of the evolution operator numerically even in the
presence of point disorder. We describe details of this
calculation in Appendix~\ref{sec:ap_b}. Here we just mention that
we discretize both space and time and write the evolution operator
as a product of transfer matrices. We take a periodic array of
$M=201$ sites in the space direction and of size $L=50$ in the
timelike direction. The particle filling factor is taken to be
approximately $0.1$, so that the ground state eigenfunction
$|G\rangle$ is the Slater determinant of the $21$ highest
eigenstates. We took the odd number of sites to have an exact
inversion symmetry around the columnar pin in the finite size
system and we took the odd number of eigenstates to avoid
complications arising from the double degeneracy of the energy
spectrum in the absence of point disorder. A columnar defect of
strength $V_0=0.1$ is placed in the central site, $x_0=101$. Point
disorder is modeled by a uniformly distributed uncorrelated random
potential on each site of the space-time lattice: $U(x,\tau)\in
[-U_0,U_0]$, so that $\Delta_0=U_0^2/3$. For the effective mass in
Eq.~(\ref{hf}) we choose $m=5$ corresponding to a hopping
amplitude $J=0.1$ in the discretized model (see
Appendix~\ref{sec:ap_b}). For each configuration of disorder we
numerically find the ground state $|G\rangle$ and the fermion
density and finally average over different realizations of point
disorder. In this way we obtain ``Friedel oscillations'' of
density for different values of $\Delta_0$. In Fig.~\ref{fig1a} we
plot a calculated density profile of vortex lines for
$U_0=10^{-3/2}$ corresponding to $\Delta_0\approx 3\times 10^{-4}$
for a particular realization of point disorder (top graph) and
after averaging over about $130\, 000$ disorder realizations
(bottom graph).
In terms of $\tilde\Delta_0$ (\ref{tilde_delta}) the chosen
strength of point disorder corresponds to
\begin{displaymath}
\tilde\Delta_0\approx {8\pi \Delta_0\over \Lambda^2}\approx 0.3,
\end{displaymath}
where we used the fact that the cutoff is $\Lambda\approx 0.1$ for
the filling factor $0.1$.

\begin{figure}[ht]
\includegraphics[width=8.3cm]{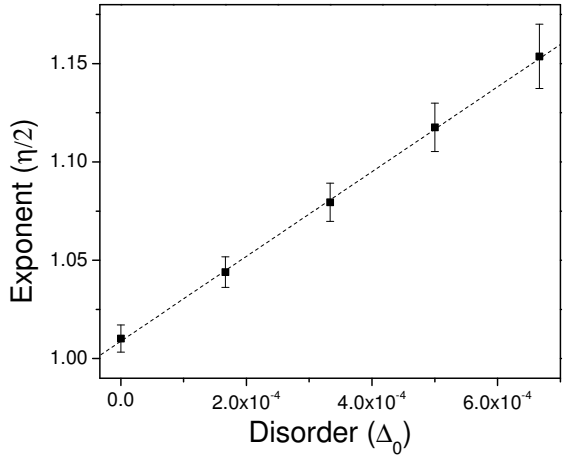}
\caption{Extracted exponent $\eta/2$ characterizing decay of
density oscillations of vortex lines near a columnar pin (see
Eq.~(\ref{36})) versus the point disorder strength $\Delta_0$. The
dashed line represents a linear fit, which agrees with the
renormalization group prediction for $g=1$ (see Eqs.~(\ref{del}),
(\ref{kap}), and (\ref{eq:32})).}
\label{fig1}
\end{figure}

Upon fitting the decay of the envelope of oscillations (dashed
line in Fig.~\ref{fig1a}) to a power law (see Eq.~(\ref{36})) for
different strengths of point disorder, we extract the exponent
$\eta/2$. The results are plotted in Fig.~\ref{fig1}.
Within the error bars the dependence of $\eta$ on $\Delta_0$ is
linear as predicted by the renoramlization group analysis for
$g=1$ (see Eqs.~(\ref{kap}) and (\ref{eq:32})).

\subsection{Boson phase correlations}

Despite the fact that dislocations can not be present in vortex
arrays (as they are equivalent to magnetic monopoles), they can be
relevant for other problems, for example directed polymer arrays
in two dimensions. The correlation function, which gives the
energy of a dislocation pair~\cite{shh} (or in the
quantum-mechanical language the boson-boson correlation
function~\cite{AHNS}) can be calculated similarly to the
density-density correlation function
\be G(x)=\langle
\overline{\mathrm e^{i[\phi(x,\tau)-\phi(0,\tau)]}\rangle}\,,
\label{g}
\ee
where the boson phase angle $\phi$ is conjugate to $d u/dx$. Upon
integrating out $\phi$ in the standard way, it is easy to show
that Eq.~(\ref{g}) can be rewritten as
\be
G(x)=\overline{\left\langle \exp\left(-\int_0^x \partial_\tau
u(x^\prime,0)dx^\prime\right)\right\rangle}.
\ee
The expression above has to be properly regularized to be cutoff
independent. In the absence of a columnar defect, the function
$G(x)$ can be straightforwardly calculated as in
Eq.~(\ref{eq:30}), yielding
\be
G(x)\propto \mathrm e^{-{1\over 2g^2}\int_0^\infty dl
[g+\kappa(l)][1-J_0(r\mathrm e^{-l})]}. \label{eq:35}
\ee
For $g\geq 1$ at large $x$ this gives
\be
G(x)\propto x^{-\eta/4g^2},
\label{gx1}
\ee
where $\eta$ is defined in Eq. (\ref{eq:32}). This reduces to the
result in the absence of point disorder when $\kappa(l)\equiv 0$:
$\zeta(x)\propto x^{-1/2g}$. For $g<1$ as $x\to\infty$ we derive
\be G(x)\propto \mathrm e^{-\epsilon^2\ln(x)^2/4g^2}. \label{gx2}
\ee
Since the columnar pin is always irrelevant in the presence of
point disorder, it will give only a perturbative correction in
$V_0$ to Eqs.~(\ref{gx1}) and (\ref{gx2}).

Note that the asymptotic behavior of $G(x)$ (Eqs.~(\ref{gx1}) and
(\ref{gx2})) is valid only for an isolated pair of bosons
(dislocations in the original problem). It can be
shown~\cite{Cardy} that under the renormalization group bosons are
relevant on both sides of the transition $g=1$. Thus, if isolated
bosons (dislocations) are permitted, they destroy the
supersolid-vortex glass transition.

\section{Response to a transverse magnetic field}
\label{sec:tilt}

Another way to determine the effect of a single columnar pin on
the flux array is to study the response of the vortices to a weak
transverse magnetic field $h$. In Refs.~[\onlinecite{AHNS}] and
[\onlinecite{ahns1}] it was argued that in the absence of point
disorder in the thermodynamic limit an {\em infinite} number of
vortices can be pinned by a single columnar defect in the limit
$h\to 0$. While for $g>1$ the fraction of pinned vortex lines goes
to zero with the sample size, for $g< 1$ almost all lines get
pinned. In the presence of point disorder, because the columnar
pin is always irrelevant, the situation is quite different. In
fact, as we show below, the number of pinned vortices $N_p(g)$
{\em has a maximum} around $g=1$.

The presence of a transverse magnetic field is manifested through
an additional term to the action in
Eq.~(\ref{act1})~\cite{hnv,hatano,ahns1}
\be
S_m=-{\pi h\over 2g}\int dxd\tau\, \partial_\tau u(x,\tau).
\ee
It is easy to see that $h$ is formally a relevant variable under
renormalization group transformation -- in fact its recursion
relation is~\cite{hnv} $dh/dl\approx h$. However, when the
columnar pin is absent the effect of the transverse magnetic field
can be eliminated by a shift of the vortex displacement field
$u(x,\tau)\to u(x,\tau)-h\tau$. Clearly, the contribution to the
action (after performing the replica trick) coming from point
disorder in Eq.~(\ref{spd}) is invariant under this
transformation. Thus, the only effect of the shift on the action
arises from the pinning term (\ref{spin}).

\subsection{Friedel oscillations}

The presence of $h$ modifies the decay of the Friedel density
oscillations around the pin~\cite{ahns1}. It is easy to see that
within linear response theory in $V_0$ the expression (\ref{34})
transforms to
\be
\overline{\langle \delta n(x)\rangle} \approx
V_0 \cos (2\pi n_0 x)\,\int_{-\infty}^\infty f(x,\tau)\cos(2\pi
h\tau) d\tau. \label{35}
\ee
In particular, for $g\geq 1$ the density modulation at large
distances and weak magnetic fields is readily obtained using
Eq.~(\ref{eq:30}):
\be
\overline{\langle\delta n(x)\rangle}\propto
V_0{\cos 2\pi n_0 x\over x^{\eta-1}} (hx)^{\eta-1\over
2}K_{\eta-1\over 2}(2\pi hx),
\label{40}
\ee
where $K_\nu (x)$ is the modified Bessel function of the second
kind. One can analyze the asymptotic behavior of Eq.~(\ref{40}) at
small and large $x$. Thus, if $h x\ll 1$ (but $x\gg 1$)
\be
\overline{\langle \delta n(x)\rangle}\propto V_0{\cos 2\pi n_0
x\over x^{\eta-1}}\biggl[1+{\sqrt{\pi}\Gamma\left({\eta\over
2}\right)(2\pi h x)^{\eta-1}\over \cos\left({\pi\eta\over
2}\right)\Gamma(\eta)\Gamma\left({\eta-1\over 2}\right)}\biggr].
\label{41}
\ee
The second term in the expression above gives a negative
correction to the result obtained in the absence of $h$. In the
opposite limit $h x\gg 1$, density fluctuations decay
exponentially with $x$:
\be
\overline{\langle \delta n(x)\rangle}\propto V_0\cos 2\pi n_0 x\,
{h^{{\eta\over 2}-1}\over x^{\eta\over 2}}\mathrm e^{-2\pi h x}.
\label{42}
\ee
These findings are consistent with the results of
Refs.~[\onlinecite{AHNS,ahns1}], where in the pure case when
$\eta=2g$. For $g<1$ the situation is more involved. Thus, for
$x\ll \mathrm e^{l_0}$ one can still use Eqs.~(\ref{41}) and
(\ref{42}) with $\eta\sim 2g$. On the other hand if $x\gg \mathrm
e^{l_0}$ the asymptotic behavior of $\overline{\langle \delta
n(x)\rangle}$ is
\be
\overline{\langle \delta n(x)\rangle}\propto \mathrm
e^{-\epsilon^2\ln^2(x)}
\ee
for a weak magnetic field, $h x\ll 1$. In the opposite limit $h
x\gg 1$ we get Eq.~(\ref{42}) with a running exponent $\eta$:
$\eta(x)\approx \epsilon^2\ln x$.

\subsection{Number of pinned vortex lines}

A much more striking effect of point disorder on the pinning
properties of a columnar defect can be observed in the number of
vortex lines prevented from tilting by the defect. This ``pinning
number'' is defined as~\cite{AHNS,ahns1}
\be
N_p(h)=n_0{\overline{\left< \int dx \left(h-{\partial
u(x,\tau)\over\partial\tau}\big|_{\tau=0}\right)\right>}\over h},
\ee
where as before $n_0$ is the mean vortex density. This definition
can be understood as follows: The numerator is proportional to the
difference between the total imaginary currents carried by the
vortices in the absence and in the presence of a columnar pin,
respectively. Dividing this current difference by the average
slope $h$ of the vortex lines in the absence of the pin gives the
effective number of vortices which do not participate in the
current flow due to the columnar defect. We identify this as the
pinning number. It is easy to see that the lowest order correction
to the current due to the pinning potential appears in second
order perturbation theory. Upon using the simple identity
\be
{\partial u(x,\tau)
\over\partial\tau}\bigg|_{\tau=0}=\lim_{\delta\tau\to 0}{\mathrm
e^{2\pi i [u(x,0)-u(x,\delta\tau)]}-1\over 2\pi i\delta\tau},
\ee
we can adopt our previous calculational technique to find
\begin{widetext}
\be
N_p\approx {n_0 V_0^2\over 4\pi h}\int\int\int d\tau d\tau^\prime
dx\, f(0,\tau-\tau^\prime)\left({\tau R(x,\tau)\over
\sqrt{x^2+\tau^2}}-{\tau^\prime R(x,\tau^\prime)\over
\sqrt{x^2+\tau^{\prime\,2}}}\right) \sin \left[2\pi h
(\tau-\tau^\prime)\right],
\ee
\end{widetext}
where
\be
R(x,\tau)=\int d l\, [g+\kappa(l)]\,\mathrm
e^{-l}J_1(\sqrt{x^2+\tau^2}\,\mathrm e^{-l}) .
\label{rx}
\ee
Note that if $|\epsilon|\ll 1$ and $\tilde\Delta_0\ll 1$ we can
neglect $\kappa(l)$ relative to $g$ in the integral to obtain
\be
R(x,\tau)\approx {1-J_0(\sqrt{x^2+\tau^2})\over
\sqrt{x^2+\tau^2}}.
\label{rx1}
\ee
After integrating over $x$ we find
\beq
&&N_p\approx {gV_0^2 n_0\over 4\pi h}\int\int d\tau d\tau^\prime
\,[\mathrm{sgn}(\tau)-\mathrm{sgn}(\tau^\prime)]\nonumber\\
&&~~~~~~~~~~~~~\times f(0,\tau-\tau^\prime) \sin 2\pi h
(\tau-\tau^\prime).
\label{Np}
\eeq
The above integral needs to be handled carefully since it is
sensitive to the order of limits. One way to deal with this is to
recall that in physical systems the integral over the imaginary
time is limited by the sample size $\tau\in[-L_\tau/2,L_\tau/2]$.
Then, in Eq.~(\ref{Np}) we can make substitutions $\tau\to
\tau+\xi/2$, $\tau^\prime\to \tau-\xi/2$ and use the identity
\be
\lim_{L_\tau\to\infty}\int_{-L_\tau/2}^{L_\tau/2} d\tau\,\left[
\mathrm{sgn}(\tau+\xi/2)-\mathrm{sgn}(\tau-\xi/2)\right]=\xi.
\ee
In this way our final expression for the number of pinned vortices
becomes
\be
N_p\approx {gV_0^2 n_0\over 4\pi h}\int d\xi\, \xi f(0,\xi)\sin
2\pi h\xi\,.
\label{np2}
\ee
{\em Case I, $g\geq 1$}. In the high-temperature phase we can use
the asymptotic behavior $f(0,\xi)\propto \xi^{-\eta}$ to obtain
\be
N_p\propto {V_0^2\over h^{3-\eta}}\,.
\label{np1}
\ee
In the absence of disorder, $\eta=2g$ and the expression above
agrees with the one obtained in Ref.~[\onlinecite{ahns1}] in the
limit $V_0\ll h$. If the sample size in either the timelike
direction $L_\tau$ or in the spacelike direction $L_x$ is finite,
then at small magnetic fields (\ref{np2}) saturates at
\be
N_p\propto V_0^2 L^{3-\eta},
\ee
where $L=\min\{L_x,L_\tau\}$. To see this we observe that if
$L_\tau$ is finite, then the integral over $\xi$ in
Eq.~(\ref{np1}) is taken within a finite interval $|\xi|\leq
L_\tau$. If $L_x$ is bounded then it is easy to show that the
correlation function $f(0,\xi)$ decays exponentially
$f(0,\xi)\propto \mathrm e^{-\xi/L_x}$ for $\xi>L_x$ and the
integral in Eq.~(\ref{np1}) is again cutoff at $\xi\approx L_x$.
Note that since with point disorder $\eta>2$, the fraction of
pinned vortices always vanishes in the thermodynamic limit
($L\to\infty$).

{\em Case II, $g<1$}. In the low-temperature vortex glass phase,
because of the faster decay of correlations with distance
\be
f(0,\xi)\propto \mathrm e^{-\epsilon^2\ln^2\xi},
\ee
the number of pinned vortices does not diverge as $h\to 0$.
Instead it saturates at
\be
N_p\propto \left({\epsilon\over \tilde\Delta_0}\right)^{1/
2\epsilon}+ \exp\left({1\over 4\epsilon^2}\right).
\label{np3}
\ee
The first term here comes from relatively short distances
$\xi\lesssim \exp(l_0)$ and the second one originates from
$\xi\gtrsim \exp(l_0)$, where $l_0$ is defined in Eq.~(\ref{l0}).
Although $N_p(h)$ remains finite as $h\to 0$ for $g<1$, note that
it becomes very large for $\tilde\Delta_0\ll \epsilon$ and
diverges exponentially at $\epsilon\to 0$ at fixed
$\tilde\Delta_0$.
\begin{figure}[ht]
\includegraphics[width=8.3cm]{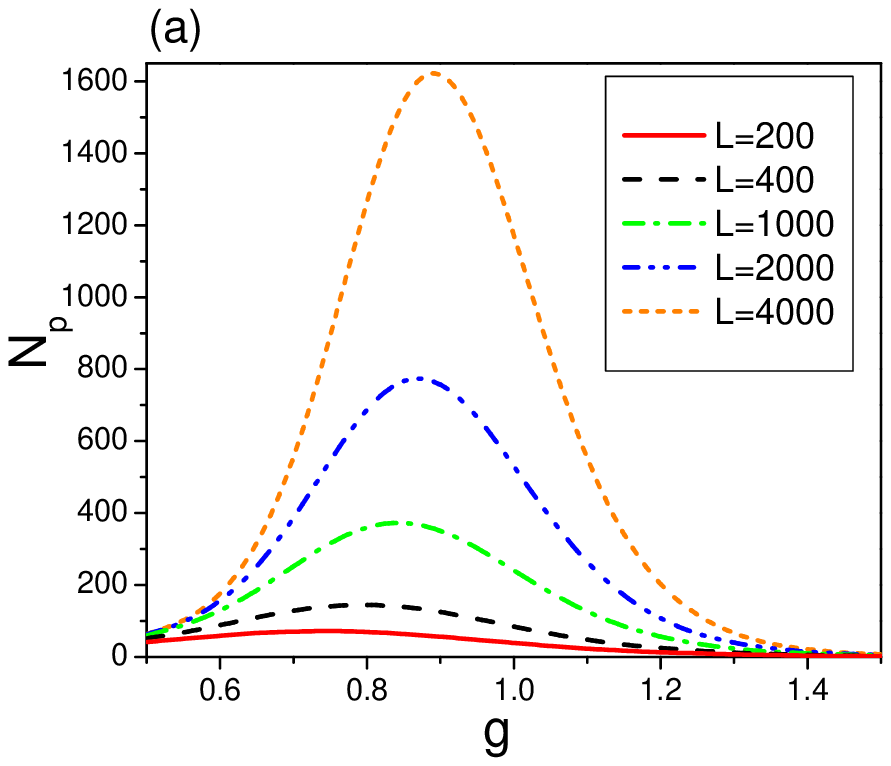}
\includegraphics[width=8.3cm]{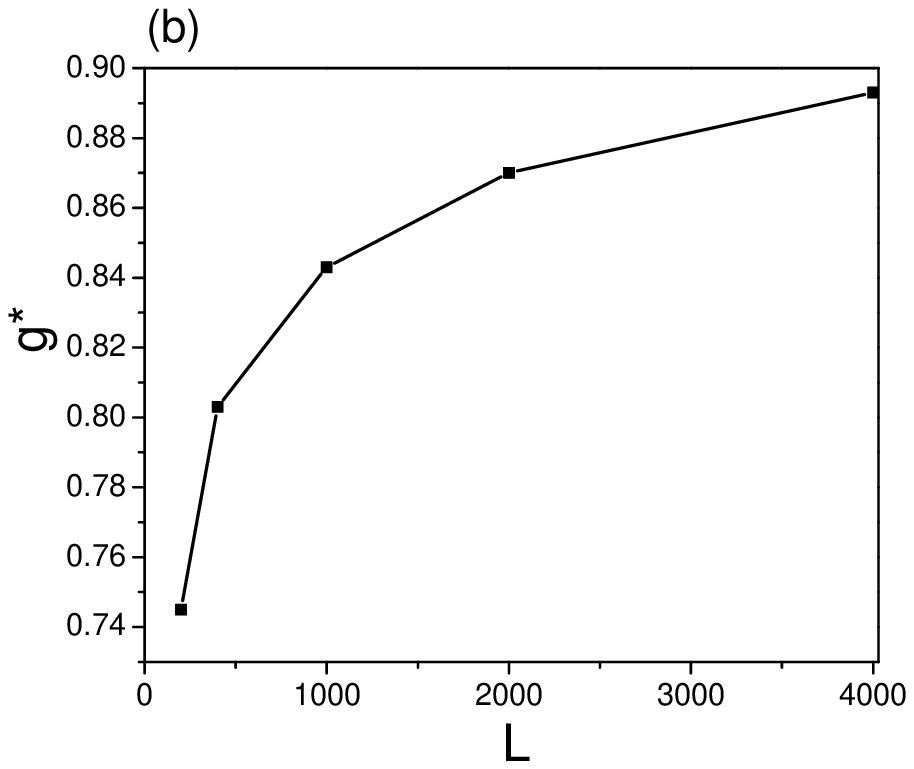}
\caption{(a) Number of pinned vortices versus the Luttinger-liquid
parameter $g$ at a fixed point disorder strength
($\tilde\Delta_0=0.2$) for different sample sizes $L$ and (b) the
position of the maximum ($g^\ast$) as a function of $L$. The value
of $g^\ast$ approaches $1$ with increasing $L$. }
\label{fig3a}
\end{figure}

An important consequence of saturation of $N_p$ at zero tilt with
the system size $L$ for $g<1$ is that as $L\to\infty$, the pinning
number will have a maximum as a function of the Luttinger-liquid
parameter at $g\approx 1$. If $L$ is not too large (or tilt is
finite) then at a given disorder strength the maximum of
$N_p(h=0)$ will occur at some $g^\ast<1$. Indeed, if $g$
approaches $1$ from below, then $N_p(h=0)$ diverges exponentially
(see Eq.~(\ref{np3})). In turn, this implies that if $L$ is finite
then the system will be insensitive to the saturating effect of
point disorder if $g$ is very close to 1 and the pinning will be
quite effective for $g\lesssim 1$ as in the case without point
disorder~\cite{AHNS}. These simple considerations agree with
numerical evaluation of $N_p$ according to Eq.~(\ref{np2}), which
are plotted in Fig.~\ref{fig3a}. The position of the maximum
($g^\ast$) versus the system size is also shown in this figure. It
clearly approaches $g^\ast=1$ as $L$ increases. In practice the
parameter $g$ can be changed varying vortex line density or
temperature~\cite{AHNS,ahns1}. Therefore measuring the dependence
of the pinning number $N_p$ on $g$ should be experimentally
feasible.

\section{Conclusions}
\label{conclusions}

In this paper we studied the effect of point disorder on one
dimensional arrays of vortex lines in the presence of a single
columnar pin. We obtained the renormalization group flow equations
for the pinning strength in the presence of point disorder and
found that the columnar pin is irrelevant at the largest length
scales (by different mechanisms) both for $g>1$ and $g<1$, which
is in contrast to the pure system where it is relevant at $g<1$.
Because the onset of relevance of point disorder also occurs at
$g=1$ we were able to quantitatively solve the RG equations for
$g$ close to 1. In particular we found that if disorder strength
is weak, then at $g<1$ the renormalized pinning potential $V$
first grows with the increasing length scale to $V_{max}\approx
V_0\sqrt{(1-g)/\tilde\Delta_0}$ and then goes to zero faster than
exponentially.

Using renormalization group analysis we calculated ``Friedel
oscillations'' of the vortex line density near the columnar pin.
In particular we showed that at $g\geq 1$ density fluctuations
decay as a power law $\delta n(x)\propto 1/x^{\eta-1}$, where the
exponent $\eta$ is shifted upwards from $\eta=2g$ due to point
disorder. If $\tilde\Delta_0\ll g-1$, then the exponent $\eta$
becomes $x$ independent only at extremely (exponentially) large
$x$. At intermediate $x$ one would observe a power law decay of
$\delta n(x)$ with an effective exponent smoothly varying with
$x$. In the case $g=1$ (which corresponds to dilute vortices with
a hard core repulsion) we mapped the problem to that of free
fermions, which we exactly solved by direct diagonalization. We
found that the decay of $\delta n(x)$ in this case agrees with the
predictions of our renormalization group analysis. For $g<1$ we
showed that the density oscillations at large distances decay
faster than a power law $\delta n(x)\propto 1/x^{(1-g)^2\ln x}$.

We found that the strongest manifestation of point disorder on
properties of a single columnar pin is the nonmonotonic dependence
of the number of vortex lines prevented by the pin from tilting
($N_p$) on $g$ in a weak transverse magnetic field. In particular,
for $g\geq 1$, this number diverges at $h\to 0$ as a power law
$N_p\propto h^{\eta-3}$, provided $\eta<3$. In finite systems this
divergence is cut off by the system size and $N_p$ saturates
$N_p(h=0)\propto L^{\eta-3}$, where $L$ is the smallest of $L_x$
and $L_\tau$. The only difference with the pure case is that
$\eta$ gets renormalized by point disorder $\eta>2g$. On the other
hand if $g<1$ then arbitrarily weak point disorder leads to
saturation of $N_p$ at $h\to 0$. As a result in large samples
$L\to \infty$, $N_p(g)$ has a maximum at $g^\ast=1$, while if $L$
is finite then the maximum occurs at some $g^\ast\lesssim 1$.

\acknowledgements

We would like to acknowledge helpful discussions with I. Affleck
and W. Hofstetter. Dr. Hofstetter particularly emphasized the
mapping to free fermions and shared his ideas on numerical
evaluation of the partition function in the presence of point
disorder. This work was supported by NSF under grants DMR-0233773,
DMR-0231631, DMR-0229243 and through the Harvard MRSCC via NSF
grant DMR-0213805.

\appendix

\section{Estimates for high temperature superconductors}
\label{sec:ap_a}

In this appendix we estimate effects of point disorder discussed
above in realistic high temperature superconductors. The point
disorder correlator $\Delta$ introduced in Eq.~(\ref{delta}) can
be estimated as~\cite{blatter, ahns1}
\be
\Delta\approx \left(\Phi_0\over 4\pi\lambda\right)^4 {\xi^3\over
T^2 wa^2} \left(j_c\over j_0\right)^{3/2}, \label{b1}
\ee
where $\Phi_0\approx 2.07\times 10^{-7}\;\rm{G\, cm}^2$ is the
flux quantum, $\lambda$ is the magnetic penetration length, $\xi$
is the superconducting coherence length, $w$ is the sample
thickness, and $a=1/n_0$ is the one-dimensional density of vortex
lines. Also $j_c$ is the critical current in the presence of point
disorder and $j_0$ is the pair-breaking current ($j_c\approx j_0
\xi^2 n_{imp}^{2/3}$, where $n_{imp}$ is the density of point
defects). The factor of $1/w$ in Eq.~(\ref{b1}) appears after
averaging over the sample thickness. We are interested at the
parameter regime where $g\approx 1$. Then using
$\tilde\Delta\approx \Delta/\Lambda^2$ (see Sec.~\ref{sec:RG}) we
get
\be
\tilde\Delta\approx\left(\Phi_0\over 4\pi\lambda\right)^4
{\xi^4\over T^2 wa} \left(j_c\over j_0\right)^{3/2}. \label{b2}
\ee
Let us take the following estimates valid for typical HTSC's:
$\lambda=150\,\mathrm{nm}$, $T=50$ K, $\xi=2$ nm, and
$j_c/j_0=0.01$. Following Ref.~[\onlinecite{bolle}] we also assume
that $w\approx\lambda$. Then Eq.~(\ref{b2}) becomes
\be
\tilde\Delta\approx 10\, {\xi/ a}. \label{b3}
\ee
Quantitative predictions made throughout this paper require that
point disorder is relatively weak: $\tilde\Delta\ll 1$. Although
we expect that arbitrarily large disorder strength renormalizes to
zero for $g>1$ and to small values $\tilde\Delta^\star\approx 1-g$
for $g\lesssim 1$ (see Eq.~(\ref{eps})), it might be necessary to
go to large length scales in order to see this effect. From
Eq.~(\ref{b3}) we see that the condition $\tilde\Delta\ll 1$ is
easily satisfied for magnetic fields below $H_{c2}$ (we recall
that in superconducting films with vortices the in-plane magnetic
field $H$ is related to the interline spacing $a$ and the film
width $w$ through $H\approx \Phi_0/aw$). On the other hand, to be
observable, $\tilde\Delta_0$ should at least be comparable to
$|1-g|$ (see, for example, Eq.~(\ref{flow1})). If the system is
tuned within $10\%$ of the vortex glass transition, i.e.
$|1-g|=0.1$, then effects of point disorder become relevant for
$a\lesssim 100\,\xi\approx 200$~nm. Such a separation between
vortex lines is comparable with the magnetic penetration length
$\lambda$ and thus should be readily accessible at magnetic fields
slightly above $H_{c1}$.

The other quantity crucial to our analysis is the Luttinger
parameter $g$. Interesting and nontrivial effects due to delicate
competition between thermal fluctuations and point disorder occur
near the vortex glass transition corresponding to $g=1$ (see, for
example, Fig.~\ref{fig3a}). In Ref.~[\onlinecite{ahns1}] it was
argued that $g$ may have a nonmonotonic dependence on the vortex
line density $n_0=1/a$. In particular, at $a\ll \lambda$ the
Luttinger liquid parameter is very small $g\ll 1$ and in the
opposite limit $a\gg\lambda$ we have $g\to 1$. It was also argued
in Ref.~[\onlinecite{ahns1}] that at least for some parameter
values, increasing the vortex line density causes $g$ to first
increase and then decrease to zero. So one might expect that $g$
crosses unity at some intermediate value of $a$ of the order of
$\lambda$, i.e., around a few hundred nanometers in typical
HTSC's. As we showed above, this is precisely what we need to
achieve point disorder strength of the order of a few percent. Any
necessary fine-tuning of $g$ and $\tilde\Delta$ can be achieved
slightly varying temperature and the in-plane magnetic field,
which determines $a$.

\section{Density oscillations of free fermions
in the presence of point disorder}
\label{sec:ap_b}

The partition function of free fermions in a potential with point
disorder (assuming for simplicity periodic boundary conditions in
the timelike direction) is given by
\be
Z=\mathrm{Tr}{\langle \Psi|  T_\tau \mathrm e^{-\int_0^{L_\tau}
d\tau\int dx \mathcal H_f(x,\tau)} |\Psi\rangle},
\ee
where Tr denotes the trace over all possible Slatter determinants
$|\Psi\rangle$ and $\mathcal H_f$ is the Hamiltonian (\ref{hf}) of
the noninteracting fermions. A slightly different matrix element
is required  for vortices with free boundary conditions and finite
$L_\tau$. However, in the limit $L_\tau\to\infty$, this detail is
irrelevant~\cite{hatano}.

\begin{figure}[ht]
\includegraphics[width=8.3cm]{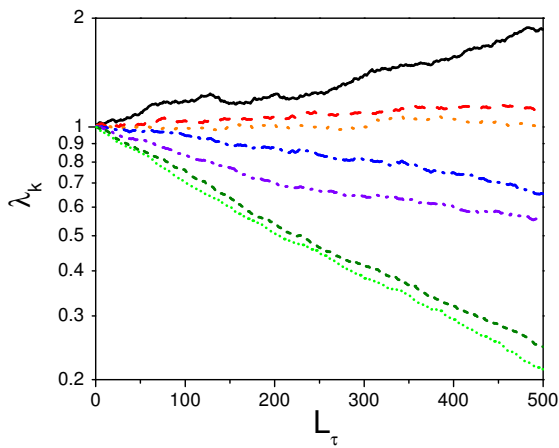}
\caption{Seven largest eigenvalues of the evolution operator
$\mathcal T$ as a function of the sample size $L_\tau$. The
parameters used in this figure are $L_x=101$, $\Delta_0\approx
0.008$, $V_0=0.02$, $J=0.1$ (see Eq.~(\ref{hf1})). For a pure
system ($\Delta_0=0$), the largest eigenvalue corresponding to a
bound state is followed by three degenerate pairs of smaller ones.
Point disorder slightly lifts degeneracy between the states, but
clearly does not affect exponential growth of the separation
between the pairs of eigenvalues with $L_\tau$.} \label{fig a1}.
\end{figure}

If the Hamiltonian were time independent, then in the zero
temperature limit (where in the quantum language temperature is
equivalent to $1/L_\tau$) the only contribution to the
$N$-particle partition function comes from the Slatter determinant
of $N$ eigenstates of $\mathcal H_f$ with lowest eigenenergies
($\varepsilon _j$). Alternatively these are the states, which form
the $N$ highest eigenstates  of the evolution operator
\be
\mathcal T=T_\tau \mathrm e^{-\int_0^{L_\tau} d\tau\int dx
\mathcal H_f(x,\tau)}. \label{s}
\ee
Note that, in a time-independent problem, nondegenerate
eigenvalues $\lambda_j$ of $\mathcal T$ become exponentially
separated from each other with increasing $L_\tau$:
\be
{\lambda_j\over \lambda_{j+1}}\propto \mathrm e^{L_\tau
(\varepsilon_{j+1}-\varepsilon_j)}.
\ee
Therefore, only the largest eigenvalues of the evolution operator
need to be considered. We are interested here, however, in the
case where the Hamiltonian explicitly depends on time due to point
disorder. In this case one can not use eigenstates of $\mathcal
H_f$, because they also depend on time. However, it is still
possible to define the spectrum of the evolution operator for a
given $L_\tau$. It seems reasonable to assume that even in the
presence of point disorder the separation between the eigenvalues
of ${\cal T}$ grows exponentially with $L_\tau$.\footnote{Because
in the disorder free case the eigenvalues of $\mathcal T$ come in
degenerate pairs, weak disorder will not necessarily strongly lift
this degeneracy. However, this issue is unimportant if we require
that in finite system the filling factor is such that these
degenerate or nearly degenerate pairs of states are either
simultaneously filled or empty.} This statement can be checked
numerically (see Fig.~\ref{fig a1}). Therefore, in the limit
$L_\tau\to\infty$ the partition function (\ref{zq1}) is dominated
the ``ground state'' $|\Psi\rangle=|G\rangle$ of the evolution
operator, i.e., by the Slatter determinant of the $N$ highest
eigenstates $g_i(x)$ of $\mathcal T$:
\be
|G\rangle=\det g_i(x_j).
\ee
Here $i,j=1,\ldots, N$ enumerate different single particle levels
and coordinates. Figure~\ref{fig a1} corresponds to relatively
strong point disorder and it is necessary to have large sample
sizes to assure the ground-state dominance. For weaker disorder it
will suffice to have smaller $L_\tau$ to find the exponential
separation of $\lambda_j$.

In order to obtain $|G\rangle$ numerically we discretize the
system both in space and time directions. In particular, instead
of the Hamiltonian given by Eq.~(\ref{hf}) we use
\be
\mathcal
H_f(\tau)\!=\!\sum_{j=0}^{M-1} \left[-J (c_{j+1}^\dagger
c_j+c_j^\dagger c_{j+1})+U_j(\tau)c_j^\dagger c_j \right]+V_0
c_0^\dagger c_0.
\label{hf1}
\ee
Here $c_j^\dagger$ and $c_j$ are fermion particle creation and
annihilation operators, respectively, and $M$ is the number of the
sites in the spacelike direction.

We use periodic boundary conditions so that $c_{j+M}\equiv c_j$.
The random potential $U_j(\tau_k)$ is taken to be uniformly
distributed on each space time point in the interval
$[-\sigma,\sigma]$:
\be
\overline{U_j(\tau_k) U_i(\tau_p)}=\Delta_0\,
\delta_{ji}\delta_{kp},
\label{sigm}
\ee
where $\Delta=\sigma^2/3$. We write the evolution operator as the
product
\be
\mathcal T=\prod_k \mathrm e^{-\mathcal H_f(\tau_k)\delta \tau}.
\ee
In order to reproduce the continuum limit (\ref{s}), we have to
take the time step $\delta\tau$ small enough to ensure that the
effects due to $[\mathcal H_f(\tau),\mathcal
H_f(\tau+\delta\tau)]\neq 0$ can be neglected. However, we do not
expect any qualitative changes in the results even if this
condition is violated. Of course, the time step $\delta\tau$ can
ultimately be taken to be $1$ by an appropriate scaling of the
couplings $J,\, U, \mathrm{and}\, V$. In a single particle space
both the Hamiltonian $\mathcal H_f$ and the evolution operator
$\mathcal T$ are $M\times M$ matrices. For a given realization of
disorder, it is straightforward to diagonalize $\mathcal T$ and
find its eigenvectors $g_i(x)$ using standard numerical methods.
Then the density profile is given by
\be
\langle \delta n_d(x_j)\rangle =\sum_{i=1}^N |g_i(x_j)|^2.
\label{nd}
\ee
The index $d$ emphasizes that this is the result for a given
realization of point disorder. After averaging $n_d(x_j)$ over
disorder realizations we obtain $\overline{\langle n(x)\rangle}$,
which is used in Fig.~\ref{fig1}.

\end{document}